\documentclass{eas}
\usepackage{graphicx}

\newcommand{\mic}{\mbox{$\,\mu$m}} 

\newcommand{\sirtf}{\mbox{\it Spitzer Space Telescope}}
\newcommand{\spitzer}{\mbox{\it Spitzer}}
\newcommand{\sof}{\mbox{\it SOFIA}}
\newcommand{\gtsimeq}{\raisebox{-0.6ex}{$\,\stackrel
        {\raisebox{-.2ex}{$\textstyle >$}}{\sim}\,$}}

\newcommand{\fion}[2]{[{#1}\,{\sc {#2}}]}

\TitreGlobal{The Physics of Evolved Stars}
\begin{document}

%%-----------------------------
%%      the top matter
%%-----------------------------

\title{The circumstellar dust of ``Born-Again'' stars}
\author{A. Evans}\address{Astrophysics Group, Keele University, Keele, Staffordshire, ST5 5BG, UK}
\author{R. D. Gehrz}\address{Minnesota Institute for Astrophysics, 
School of Physics and Astronomy, 116 Church Street, S.E., 
University of Minnesota, Minneapolis, Minnesota 55455, USA}
\author{L. A. Helton}\address{USRA-SOFIA Science Center, NASA Ames Research Center, Moffett Field, CA 94035, USA}
\author{C. E. Woodward}\sameaddress{2}

\begin{abstract}
We describe the evolution of the carbon dust shells around Very Late Thermal Pulse (VLTP)
objects as seen at infrared wavelengths. This includes a 20-year overview of the evolution
of the dust around Sakurai's object (to which Olivier made a seminal contribution) and
FG~Sge. VLTPs may occur during the endpoint of as many as 25\% of solar mass stars, and
may therefore provide a glimpse of the possible fate of the Sun.
\end{abstract}
\maketitle
%%-----------------------------
%%      your text
%%-----------------------------
\section{Introduction}

It is well-known that the fate of a star after it has
evolved away from the Main Sequence (MS) depends on its mass. The accepted
scenario for the post-MS evolution of low to inter\-mediate mass stars is that,
following the helium flash, burnout of He occurs in the core on the horizontal
branch. After evolution up the Asymptotic Giant Branch, the star sheds its
outer envelope, which is illuminated as a planetary nebula (PN) by the still-hot stellar core.

However, in as many as 20\% of cases (Bl\"ocker \cite{blocker}) the star,
as it evolves towards the white dwarf (WD) region of the HR diagram, re-ignites a
residual helium shell in a VLTP and retraces its evolutionary track to the right
to become a born again red giant (BAG; see Herwig \cite{herwig} and references therein). 
The final evolution to a WD is predicted to take roughly a few centuries, thus 
representing a very rapid (and hence seldom seen) phase of stellar evolution. A
very small number of stars is known to have undergone VLTPs, e.g. Sakurai's
Object (V4334~Sgr), V605~Aql and FG~Sge. All are C-rich, H-poor,
have extensive dust shells, and lie at the centre of a PN. Observations of BAGs
in the past $\sim20$~years have shown that the rate at which the final evolution
occurs has been seriously {\em underestimated.}

\vspace{-2mm}

\section{Observations}

We have a long-standing programme of observations of BAGs at infrared (IR)
wavelengths, some extending at least over a couple of decades. These included
ground-based observations (e.g. Woodward \etal, \cite{woodward};
Tyne \etal, \cite{tyne}; Gehrz \etal, \cite{gehrz-fgsge}) and observations using the \sirtf\ Infrared Spectograph
(e.g. Evans \etal\ \cite{evans-sak}) and the FORCAST grism spectrometer on the
Stratospheric Observatory for Infrared Astronomy (\sof).

\vspace{-2mm}

\subsection{FG Sge}
FG~Sge underwent its VLTP in 1880.
The 5--36\mic\ spectrum as observed by \spitzer\ and \sof\ is shown in Fig.~\ref{bags}.
There is clearly a dust continuum, at a temperature $\sim650$~K. This is considerably
lower than that determined by either Woodward \etal\ (\cite{woodward}) or
Gehrz \etal\ (\cite{gehrz-fgsge}), who found that there
was a major dust-formation event sometime between 1983 and 1992. They also found that,
over the period 1992--2001, the dust temperature was consistently within $\pm200$~K of 1000~K,
indicating that mass-loss and dust production was continuing.

More recent \spitzer\ data show that, sometime between 1992 and 2006
(the date of the first \spitzer\ observation) mass loss had ceased and that the dust
shell had started cooling. Of interest is the difference between the underlying 650~K blackbody and
the \spitzer\ IRS spectrum, shown in Fig.~\ref{bags}. There is clear evidence for
hydrocarbon ``Unidentified InfraRed'' (UIR) emission,
and absorption by acetylene (C$_2$H$_2$;  13.688\mic);
this demonstrates that the mass lost in the 1992--2001 episode
(Gehrz \etal, \cite{gehrz-fgsge}) was carbon-rich.
Also highlighted are two weak emission features at 33.45\mic\ and 34.75\mic\ (see below).
The \sof\ FORCAST observation, obtained in 2014, show that the dust has further cooled
(see Fig.~\ref{bags}).

\vspace{-2mm}

\subsection{Sakurai's Object}
The VLTP of Sakurai's Object occurred in 1996. Thorough analyses of the IR
development and the implications for the evolution of the (carbon) dust produced
are given by Tyne \etal\ (\cite{tyne}) and Hinkle \& Joyce (\cite{hinkle}).
This object also showed C$_2$H$_2$ in absorption, as well as HCN and a range of
polyynes (HC$_n$N; Evans \etal\ \cite{evans-sak}). 
The ejection of an optically thick carbon dust shell commenced in late 1997
(Evans \etal\ \cite{evans-sak}; Hinkle \& Joyce \cite{hinkle}); following the dust
ejection event the central star has been completely ($V\gtsimeq24$) obscured,
and has only recently (2015 February) become visible at $V\simeq17$.
Exquisite observations by Chesneau \etal\ (\cite{chesneau}) with the mid-IR interferometer
MIDI/VLTI showed that the dust was confined to a disc.

\begin{figure}
\includegraphics[width=4.5cm,angle=-90]{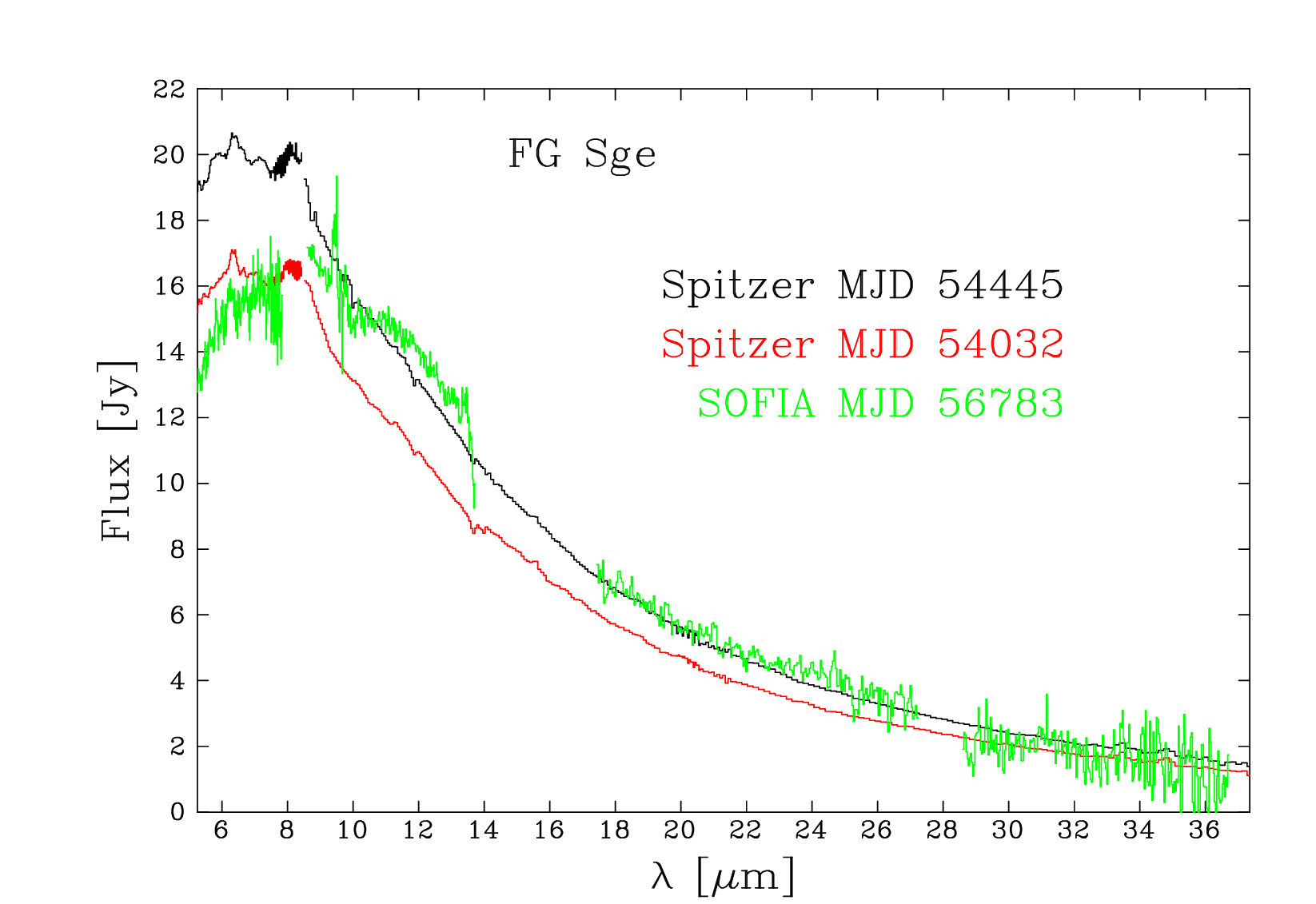}
\includegraphics[width=4.5cm,angle=-90]{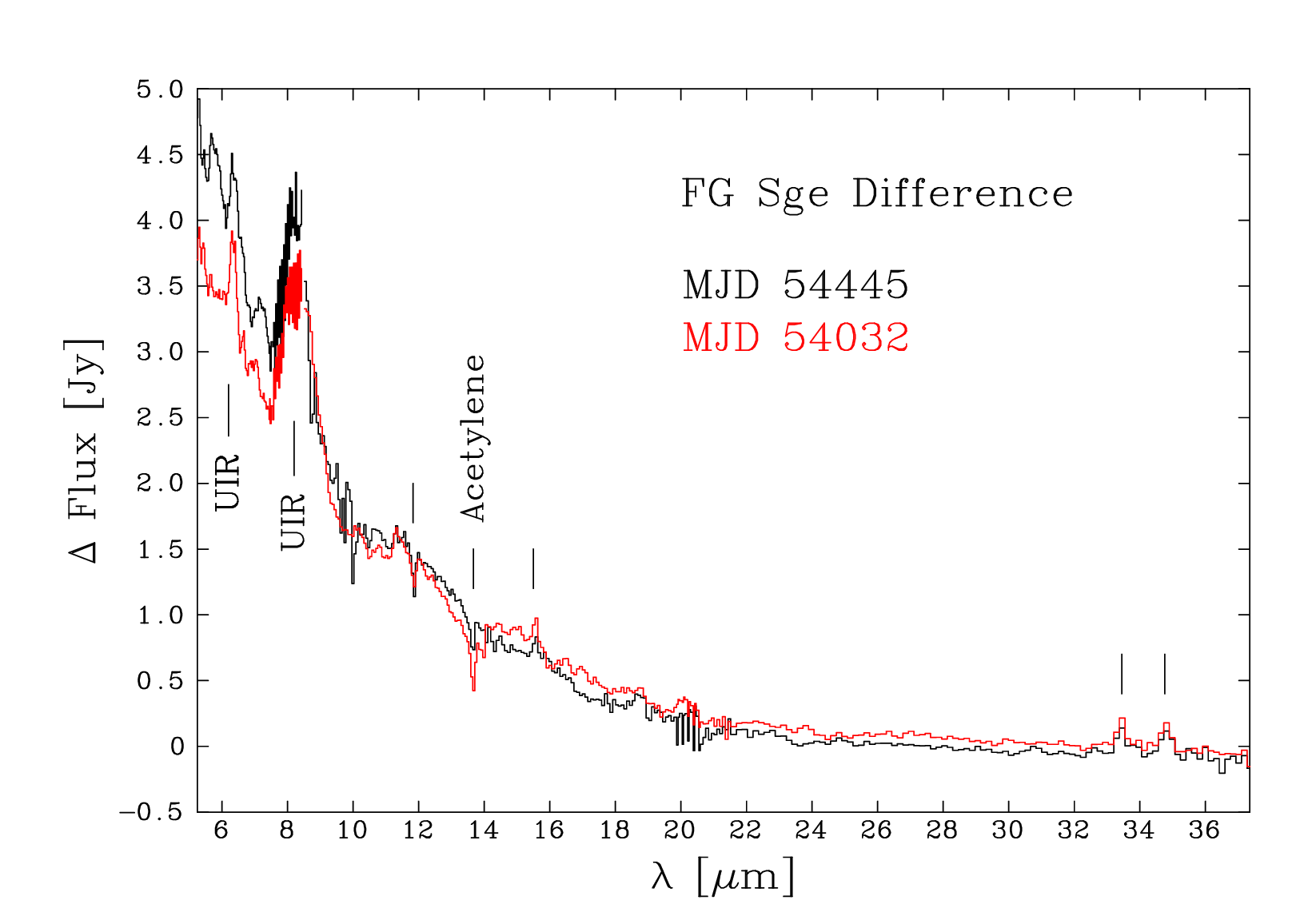}
\includegraphics[width=4.5cm,angle=-90]{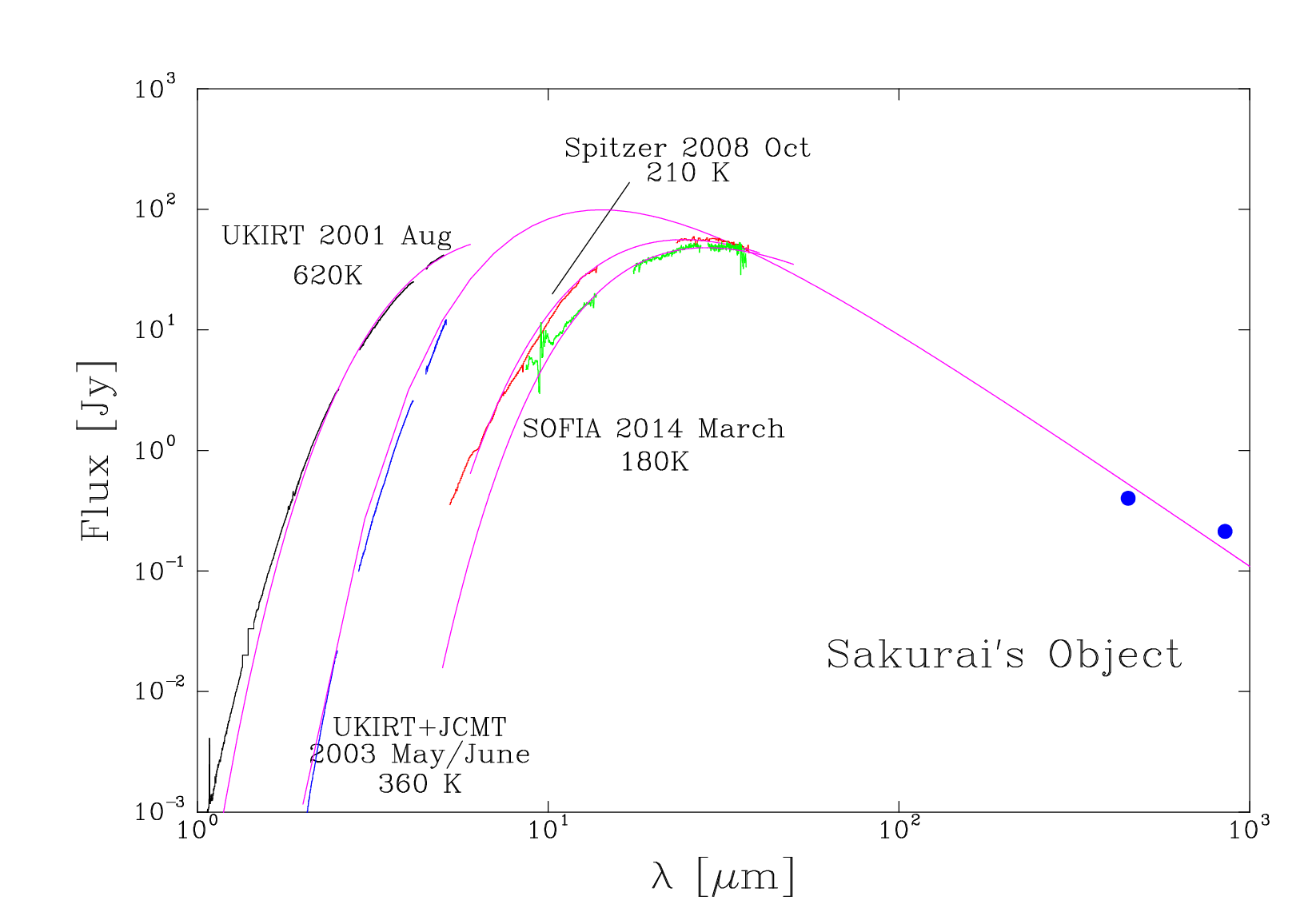}
\includegraphics[width=4.5cm,angle=-90]{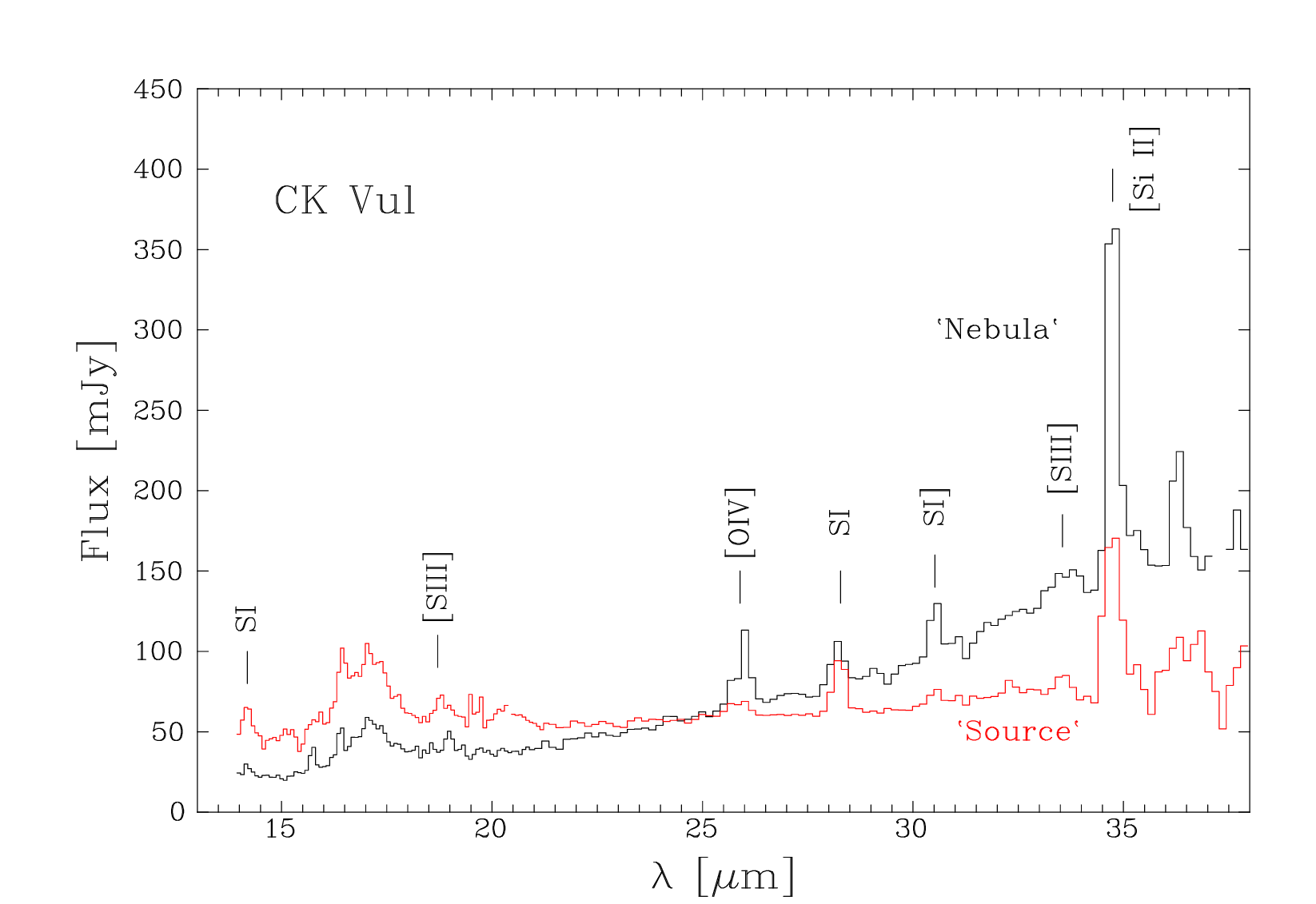}
\caption{{\bf Top left:} The \spitzer\ IRS and \sof\ FORCAST spectrum of FG~Sge.
{\bf Top right:} Difference between the \spitzer\ IRS data and a 650~K blackbody,
highlighting UIR features.
{\bf Bottom left:} Cooling of the dust shell of Sakurai's Object between 2001 and 2014 as
seen by \spitzer\ and \sof.
{\bf Bottom right:} Difference between the \spitzer\ IRS data and a 170~K blackbody for
CK~Vul, highlighting the UIR features.
\label{bags}}
\end{figure}

The cooling of the dust shell from 620~K to 180~K, as observed from the ground (UKIRT+JCMT) 
and from space- and air-borne (\spitzer, \sof) observations,
is shown in Fig.~\ref{bags}. The HCN isotopologues in the \spitzer\ IRS spectra
show that the $^{12}$C/$^{13}$C ratio is $\simeq3.2$, consistent with that
deduced from the fundamental and first overtone CO absorption (Eyres \etal, \cite{eyres};
Pavlenko \etal, \cite{pavlenko}), and with its VLTP status.

\vspace{-2mm}

\subsection{CK Vul}
The nature of CK~Vul is unclear. Once thought to be the oldest ``old nova''
(Shara \etal, \cite{shara}) -- it erupted in 1670 -- it has also been suggested that it is a BAG (Evans \etal,
\cite{evans-ck} and references therein), a
diffusion-induced nova (Miller Bertolami \etal, \cite{mm3b}), and a stellar merger
(Kaminski \etal, \cite{kaminski}).

The \spitzer\ IRS spectrum, of both the central object and the inner nebula,
shows a number of UIR features on a 170~K blackbody continuum;
the difference between the IRS spectrum of the central object,
and a 170~K blackbody is shown in Fig.~\ref{bags}.
There are a number of ionic features (e.g. \fion{O}{iv} at 25.89\mic) and we again see UIR
features that are completely different from those commonly
seen in novae (Helton \etal, \cite{helton}).

We also note the two features at 33.45\mic\ and 34.75\mic, also present in FG~Sge (see above).
They are labelled as being due to \fion{S}{iii} and \fion{Si}{ii} in Fig.~\ref{bags} but
they are extraordinarily strong in CK~Vul compared to their strength in FG~Sge. There
seems to be no reasonable way in the case of CK~Vul to account for these features in
terms of ionic emission only. It may be that they arise either in entirely
different environments in the two objects, or in different carriers.

\vspace{-2mm}

\section{Conclusion}

Observations of the BAG phenomenon at mid-to-far IR (5--38\mic) wavelengths provide
a wealth of information (e.g. about the mineralogy of the dust, far IR fine
structure lines) -- and hence about the circumstellar environment of BAGs --
that can be obtained by no other means.
Observations of BAGs using \sof\ are continuing and will be reported elsewhere.

\vspace{-2mm}


\begin{thebibliography}{99}
\bibitem[2001]{blocker} Bl\"ocker, T., 2001, Ap\&SpSci, 275, 1
\bibitem[2009]{chesneau} Chesneau, O., \etal, 2009, A\&A, 493, L17
\bibitem[2006]{evans-ck} Evans, A., \etal, 2002, MNRAS, 332, L35
\bibitem[2006]{evans-sak} Evans, A., \etal, 2006, MNRAS, 373, L75
\bibitem[2004]{eyres} Eyres, S. P. S, \etal, 2004, MNRAS, 350, L9
\bibitem[2005]{gehrz-fgsge} Gehrz, R. D., \etal, 2005, ApJ, 623, 1105
\bibitem[2011]{helton} Helton, L. A.,  \etal, 2011, in
        {\it PAHs and the Universe}, EAS Publ. Series, Vol. 46, 407
\bibitem[2005]{herwig} Herwig, F., 2005, ARA\&A, 43, 435
\bibitem[2014]{hinkle} Hinkle, K. H., \& Joyce, R. R., 2014, ApJ, 785, 146
\bibitem[2011]{mm3b} Miller Bertolami, M. M., \etal, 2011, MNRAS, 415, 1396
\bibitem[2004]{pavlenko} Pavlenko, Ya. V., \etal, 2004, A\&A, 417, L39;
\bibitem[2015]{kaminski} Kaminski, T., \etal, 2015, Nature, 520, 322
\bibitem[1985]{shara} Shara, M. M., \etal, 1985, ApJ, 294, 271
\bibitem[2002]{tyne} Tyne, V. H., \etal, 2002, MNRAS, 334, 875
\bibitem[1993]{woodward} Woodward, C. E., \etal, 1993, ApJ, 408, L37
\end{thebibliography}
\end{document}